\documentclass[aps,pre,preprint,floats]{revtex4}
\usepackage{graphics}
\usepackage[dvipdfm]{graphicx}
\usepackage{epsfig}
\usepackage{amsmath}
\begin{document}
\title{Statistical Approach to Fractal-Structured Physico-Chemical Systems: 
Analysis of non-Fickian diffusion} 
\author{Aurea R. Vasconcellos}
\affiliation{Instituto de F\'{\i}sica ``Gleb Wataghin''
Universidade Estadual de Campinas, Unicamp
13083-970 Campinas, S\~ao Paulo, Brazil}
\author{J. Galv\~ao Ramos}
\affiliation{Instituto de F\'{\i}sica ``Gleb Wataghin''
Universidade Estadual de Campinas, Unicamp
13083-970 Campinas, S\~ao Paulo, Brazil}
\author{A. Gorenstein}
\affiliation{Instituto de F\'{\i}sica ``Gleb Wataghin''
Universidade Estadual de Campinas, Unicamp
13083-970 Campinas, S\~ao Paulo, Brazil}
\author{M.U. Kleinke}
\affiliation{Instituto de F\'{\i}sica ``Gleb Wataghin''
Universidade Estadual de Campinas, Unicamp
13083-970 Campinas, S\~ao Paulo, Brazil}
\author{T. G. Souza Cruz}
\affiliation{Instituto de F\'{\i}sica ``Gleb Wataghin''
Universidade Estadual de Campinas, Unicamp
13083-970 Campinas, S\~ao Paulo, Brazil}
\author{Roberto Luzzi}
\affiliation{Instituto de F\'{\i}sica ``Gleb Wataghin''
Universidade Estadual de Campinas, Unicamp
13083-970 Campinas, S\~ao Paulo, Brazil}
\date{\today}
\begin{abstract} 
Competing styles in Statistical Mechanics have been introduced to investigate 
physico-chemical systems displaying complex structures, when one faces 
difficulties to handle the standard formalism in the well established 
Boltzmann-Gibbs statistics. After a brief description of the question, we 
consider the particular case of Renyi statistics whose use is illustrated in 
a study of the question of the ''anomalous'' (non-Fickian) diffusion that it 
is involved in experiments of cyclic voltammetry in electro-physical chemistry. 
In them one is dealing with the fractal-like structure of the thin film 
morphology present in eletrodes in microbatteries which leads to 
fractional-power laws for describing voltammetry measurements and in the 
determination of the interphase width derived using atomic force microscopy. 
The fractional-powers associated to these quantities are related to each other 
and to the statistical fractal dimension, and can be expressed in terms of a 
power index, on which depends Renyi's statistical mechanics. It is clarified 
the important fact that this index, which is limited to a given interval, 
provides a measure of the microroughness of the electrode surface, and is 
related to the dynamics involved, the non-equilibrium thermodynamic state of 
the system, and to the experimental protocol.  
\end{abstract}
\maketitle 		
%
%
\section{Introduction} 
A couple of decades ago Montroll and Shlesinger \cite{1} stated that in the 
world of investigation of {\it complex phenomena that requires statistical 
modeling and interpretation, several competing styles have been emerging}, 
each with its own champions. Lately a good amount of efforts have been devoted 
to the topic. What is at play consists in that in the study of certain 
physico-chemical systems we may face difficulties when handling situations 
involving fractal-like structures, correlations (spatial and temporal) with 
some type of scaling, turbulent and chaotic motion, small size (nanometer 
scale) systems with eventually a low number of degrees of freedom and 
complicate boundary conditions, and so on. The interest has been nowadays 
enhanced 
as a result that such situations are present in electronics, opto-electronics, 
etc, devices of the present day point-first technologies, on which is so 
dependent our society, also in disordered systems, polymeric solutions, ion 
conducting glasses, microbatteries, and others. Most cases involve, what in 
the literature on the subject are called "anomalous" physical properties ( let 
them be transport, optical, hydrodynamical, etc. ones). As a rule what is 
noticed is that the observable experimental data can be theoretically 
adjusted by means of fractional-power laws in place of the standard ones. 
We can mention one type, related to the one presented here, consisting into 
"anomalous" (non-Fickian) diffusion in micelles \cite{2,3,4,5}, where the 
analysis of data used to be done introducing the so-called L\'evy statistics. 
In this paper we present a closed statistical theory of a large scope in 
place of such approach.
   
In these situations the difficulties consist, as a rule, in that the 
researcher is impaired, for one reason or other, to satisfy Fisher's criteria 
of efficiency and sufficiency \cite{6} in the standard formalism to implement 
the conventional, well established, 
physically and logically sound Boltzmann-Gibbs statistics, meaning an 
impairement to correctly include the presence of large fluctuations (and 
eventually higher-order variances) and to account for the relevant and proper 
characteristics of the system, respectively. To contour these difficulties, 
and to be able to make predictions -- providing and understanding, even 
partial, of the physico-chemical properties of the system of interest, for 
example, in analyzing technological characteristic of a device -- there has 
been developed efforts along the line of introducing heterotypical statistics 
for providing an auxiliary method for fulfilling such purposes.

Among existing approaches it can be mentioned, what we call Generalized 
Statistical Mechanics as used by P.T. Landsberg showing that functional 
properties of the (informational) entropy (which is a generating functional 
of probability distributions) provides different types of 
thermodynamics, and rise the question of how to select a "proper" one, 
that is, are some better than others? \cite{7}; Superstatistics as used by 
C. Beck and E.G.D. Cohen for nonequilibrium systems with complex dynamics in 
stationary states with large fluctuations on long-time scales\cite{8, 9}.
	
So-called nonextensive statistics based on Havrda-Charvat statistics \cite{10} 
applied to a number of cases are described in the Conference Proceedings of 
Ref.[11]. Renyi statistics \cite{12,13} has been introduced in the 
scientific literature \cite{14}, with T. Arimitzu and P. Jizba \cite{15,16} 
recently introducing an extensive analysis in a paper intitled ''The world 
according to Renyi; Kappa (sometimes called Deformational) statistics -- 
a particular case of Sharma-Mittal statistics \cite{17} -- 
was used in problems of plasma physics in celestial mechanics \cite{18} 
and by G. Kanadiakis in the case of relativistic systems\cite{19}.  
	  
The introduction of these alternative statistics resides on the fact, already 
stated, that the well established mechanical-statistical treatment via 
Boltzmann-Gibbs ensemble formalism, let it be around equilibrium  (as in 
the usual response function theory) or for systems arbitrarily away from 
equilibrium \cite{20}, has its application impaired because of the 
complications set forth by the presence of spatial correlations, resulting 
for example in the illustration here presented from spatially varying 
microroughened boundaries showing a fractal structure. Such fractal structure 
introduces a degree of fuzziness in data and information, which consists in 
the lack of information on the space 
correlations imposed by the fractal-like microroughness of the electrode 
surface. Hence, as noticed, it is not satisfied Fisher's criteria of 
efficiency and sufficiency in the characterization of the system, that is, 
the chosen description does not summarize the whole relevant information for 
the situation in hands. Hence the use of formulations based on unconventional 
statistics \cite{22} to perform calculations of average values and response 
functions in which the accompanying heterotypical probability distributions 
preferentially weight the probabilities of the states that contribute to the 
dynamics involved \cite{23,24}. We restate that the use of unconventional 
(index dependent)  heterotypical statistical mechanics for systems 
displaying correlation effects of one kind or other, is a circumventing way 
for overcoming the difficulty to assess such characteristics that should be 
incorporated in the conventional formulation within the well established 
Boltzmann-Gibbs theory. 

We consider here Renyi statistics \cite{12,13} applied in statistical 
mechanics \cite{15,16,23,24}, illustrating its use and utility in the case 
of a particular experiment, namely, observation and measurements using cyclic 
voltammentry in the study of fractal-structure electrodes in microbatteries. 
First, we begin again noticing that the  advanced technologies that keep 
evolving 
nowadays present challenges for the associated physics which describes the 
phenomena that are in them involved thus, requiring a careful use of 
the accompanying theoretical physics to face the difficulties associated 
with the handling of these new situations. They involve aspects of the 
physics of soft-matter and disordered systems, complex structured systems, 
the already mentioned small systems, and others, as well as the question of 
functioning in far-from-equilibrium conditions and in the presence of 
ultrafast (pico- and femto-second time scales) relaxation and pumping effects. 
We are dealing in all these cases with condensed matter and then, we recall, 
it is inevitable for their study at the physico-chemical level to resort to 
statistical mechanics -- in a sense it can be said that the physics of 
condensed matter is statistical  mechanics by antonomasia. We consider here 
a particular situation which provides an excellent illustration of what can 
be  called {\it Unconventional Statistical Mechanics} at work. We begin noticing that 
as a consequence of the nowadays large interest associated to the 
technological applications in displays, electrochromic windows and 
microbatteries  (see, for example, Ref.[25]), the study of growth, annealing 
and surface  morphology of thin film depositions has acquired particular 
relevance. These kind of systems involve the presence of  microroughened 
surface boundaries in a geometrically constrained region, when fractal 
characteristics can be expected to greatly influence the physical 
properties \cite{26}. It can be recalled that fractals display 
self-similarity, but the latter is said to come in two flavors: exact and 
statistical, the latter being the one with which we are dealing here. 
Moreover, fractals are 
characterized in terms of their, in general, fractional ''dimension'' -- 
as we shall see in the present case -- which is a ''measure'' of the 
complexity of the structure.

Hence, we do have that the dynamics or hydrodynamics involved in the 
functioning of such devices can be expected to be governed by some type or 
other of scaling laws \cite{27,28,29}. Such characteristics have been 
experimentally evidenced and the scaling laws determined by researchers who 
resorted to the use of several experimental techniques, as atomic force 
microscopy \cite{30} which allows to obtain the detailed topography of the 
surface --  and then measuring the statistical fractal dimension --  and cyclic 
voltammetry \cite{31} which is an electrochemical technique used for the 
study of several phenomena, also allowing for the characterization of the 
fractal properties (scaling laws) of the system. Particularly, through 
cyclic voltammetry it can be  evidenced the property of the so-called 
"anomalous" (more properly called non-Fickian) diffusion in these systems. 
It is our aim to present here a theory 
appropriate to deal with these kind of situations. For dealing with transport 
processes in electrochemical cells we introduce two different aspects of 
recent thermo-hydrodynamic theory: one is Informational Statistical 
Thermodynamics (IST for short \cite{32,33}, which provides  statistical 
foundations to Extended Irreversible Thermodynamics \cite{34}), and a 
non-equilibrium ensemble formalism \cite{21}. However, for the kind of 
experiments to be analyzed one needs, as noticed,  to resort to a modified 
form of the standard formalism in the original and well established 
Boltzmann-Gibbs foundations of statistical mechanics. Originally, the relation 
between irreversible thermodynamics and transport laws, as Fick and Fourier 
laws, were well established -- and kinetic and mechanical statistical 
foundations given -- in the framework of classical (or Onsagerian) 
irreversible thermodynamics. But, it faces limitations since the theory is 
restricted to weak amplitudes of movement, with a linear relationship between 
fluxes (currents) and thermodynamic forces, smooth variation in space and 
time, locality in space and instantaneously in time (i.e. neglecting space 
and time correlations) and weak fluctuations, and then does not cover the 
case of fractal-structured systems. To remove these restrictions appeared a 
variety of approaches, a most convenient one is the already mentioned 
Extended Irreversible Thermodynamics and its statistical counterpart 
Informational Statistical Thermodynamics, and for a practical handling of 
fractal and nanometric systems one can resort to unconventional statistics. 
IST is founded on a non-equilibrium ensemble formalism which provides a 
generalized nonequilibrium grand-canonical ensemble \cite{21,35,36}. 
On the other hand, unconventional statistics applies to 
systems displaying long-range interactions, persistent memory, evolution in a 
fractal space, etc~\cite{23,24}. It has been applied to some kind of studies 
of ''anomalous'' diffusion, as for example in Refs.[37-39], and we reconsider 
the question here, specified to the situation in fractal-structured 
electrodes of microbatteries leading to "anomalous" results in experiments 
involving cyclic voltammetry.
\section{Theoretical Analysis}
Within the variational approach to Statistical Mechanics \cite{21} the use 
of unconventional (index-dependent) informational 
entropies (better called measure of uncertainty of information, and not to be 
confused with the physical entropy which is a property of the thermodynamic 
state of physical systems, while the informational one is a property of any 
probability distribution \cite{23,24,40}) \cite{22} leads to the construction 
of alternative Statistical Mechanics as described in 
Refs.[23,24]. Basically the process consists of two steps: (1)The choice of an 
{\it heterotypical probability distribution} \cite{22,23,24}, say $\varrho_{h}$, 
and (2) the use of an {\it escort probability} \cite{41} of order  $\gamma$ 
in terms of this heterotypical distribution, say 
${\cal D}_{\gamma}\{\varrho_ {h}\}$, namely, in the calculation of average 
values it is used the definition
\begin{eqnarray}
\left\langle \hat{A} \right\rangle = Tr \left\{ \hat{A} \, 
{\mathcal D}_{\gamma} \left\{\varrho_{h} \right\} \right\} \, ,
\label{e1}
\end{eqnarray}
where the escort probability of order $\gamma$ is given by
\begin{eqnarray}
{\mathcal D}_{\gamma} \left\{ \varrho_{h} \right\} = 
\frac{\varrho_{h}^{\gamma}}{Tr \left\{ \varrho_{h}^{\gamma} \right\}} \, ,
\label{e2}
\end{eqnarray}
where $\gamma$ is a real positive number (see for example Ref.[41], where the 
name escort probability is used, and the concept generalized, and also in  
Chapter IX, pp. 569 \emph{et seq.} of Ref.[13]). An in depth presentation and 
discussion is given elsewhere  \cite{23,24} , here suffices to say that what 
the heterotypical distribution does is providing a weighting of the 
probabilities of the states involved, and the escort probability accounts for 
the influence of correlations and higher-order variances. 

According to Beck and Sch$\ddot{o}$gl \cite{41}, the use of the escort 
probability 
introduces an increase in information (by incorporating self-consistently 
correlations and variances) in the sense that, for the case of index 
$\alpha$-dependent Renyi distribution $ \varrho_{\alpha}$ (see below)
 \begin{eqnarray}
( 1 - \alpha )^{2} \, \frac{\partial I_{\alpha}}{\partial \alpha \hfill} = 
Tr \left \{ {\cal D}_{\alpha} \left\{ \varrho_{\alpha} \right\} 
\left[ 
\ln \left( {\cal D}_{\alpha} \left\{ \varrho_{\alpha} \right\} \right) - 
\ln \left( \varrho_{\alpha} \right) \right] \right\} \, ,
\label{e3}
\end{eqnarray}
where $I_{\alpha}$ is the quantity of information (the negative of Renyi's 
informational entropy in terms of ${\varrho_ {\alpha}}$), and the 
right-hand side is interpreted as the information gain when using the escort 
probability ${\cal D}_{\alpha}$ built in terms of the original one 
${\varrho_ {\alpha}}$. 

Let us now proceed to deal with the problem on hands, when we make use of
 an unconventional statistics in Renyi approach in order to provide a 
statistical-thermodynamic treatment of the "anomalous" diffusion law proposed 
to account for the experimental results obtained using cyclic voltammetry in 
the system of fractal-structured electrodes in microbatteries. It involves the 
process of motion of charges in the electrolyte bordered by a 
fractal-structured thin film on the electrode, motion which is usually 
considered that can be described as a process of diffusion governed by Fick 
law \cite{31}. But the use of standard Fick diffusion equation does not 
properly describe the experimental results, which show some kind of power 
law (see below) differing from the expected one. This is a consequence that 
in a constrained (nanometric scaled) fractal-like geometry, Fickian-diffusion 
does not apply because are not satisfied the quite restrictive conditions 
mentioned in the Introduction. A proper description using the conventional 
approach in statistical mechanics for providing foundations for hydrodynamics 
requires the use of a generalized higher-order thermo-hydrodynamics 
\cite{42,43,44}, that is, the description including as basic macrovariables 
the density of energy, the density of particles, $h({\mathbf r},t)$ and 
$n({\mathbf r},t)$, their fluxes of all orders, 
${\mathbf I}_{h}({\mathbf r},t)$, ${\mathbf I}_{n}({\mathbf r},t)$, 
${\mathbf I}_{n}^{[r]}({\mathbf r},t)$ with $r = 2, 3, ...$ indicating the 
order of the flux and its tensorial rank; all the quantities are defined in 
Refs.[21] and [42]. To resort to this higher order hydrodynamics (which 
provides at the microscopic, i.e., statistical mechanical level, 
generalizations of super-Burnett and super-Gruyer-Krumhansl equations (see 
Refs.[45] and [38]), even in the linear aproximation and neglecting 
fluctuations is a formidable task in the present case with spatially varying 
boundary conditions (in a nanometric scale and with a complex structure). 
Moreover, even though a contraction in the choice of the 
basic set of macrovariables can be introduced \cite{21,45}, it is difficult 
to establish the order of the truncation. Therefore, what can be attempted in 
order to obtain a description of some properties of the system, is to 
introduce an unconventional statistics, leading to an unconventional 
hydrodynamics, in terms of a low-order truncation, in which one keeps only 
the variables
\begin{eqnarray}
\Big\{ h({\mathbf r},t), n({\mathbf r},t), 
{\mathbf I}_{n}({\mathbf r},t) \Big\} \, ,
\label{e4}
\end{eqnarray}
that is, the energy density, the density of particles and the first-order 
flux (current). In that way we are led to obtain an "anomalous" diffusion 
equation depending on a fractional-power which is determined by the dynamics, 
the fractality, the geometry 
and dimensions, and the thermodynamic state of the system (the unconventional 
generalized higher-order thermo-hydrodynamics is presented elsewhere 
\cite{47,48}). Let us proceed to derive their equations of evolution using 
Renyi approach to unconventional statistics \cite{23,24}. For that purpose, 
first we separate the Hamiltonian in the form 
\begin{eqnarray}
\hat{H} = \hat{H}_{0} + \hat{H}^{\prime} \, ,
\label{e5}
\end{eqnarray}
where $\hat{H}_{0}$ is the kinetic energy and $\hat{H}^{\prime}$ contains all 
the interactions that are present in the system. Moreover, in a classical 
approach, the basic dynamic variables, whose average values are those of 
Eq.(\ref{e4}), are 
\begin{eqnarray}
\hat{h}({\mathbf r}|\Gamma) = \int d^{3}p \, \frac{p^{2}}{2m} \, 
\hat{n}_{1}({\mathbf r},{\mathbf p}|\Gamma) \, ,
\label{e6}
\end{eqnarray}
\begin{eqnarray}
\hat{n}({\mathbf r}|\Gamma) = \int d^{3}p \, 
\hat{n}_{1}({\mathbf r},{\mathbf p}|\Gamma) \, ,
\label{e7}
\end{eqnarray}
\begin{eqnarray}
\hat{{\mathbf I}}_{n}({\mathbf r}|\Gamma) = \int d^{3}p \, 
\frac{{\mathbf p}}{m} \, \hat{n}_{1}({\mathbf r},{\mathbf p}|\Gamma) \, ,
\label{e8}
\end{eqnarray}
where $m$ is the mass of the particles, ${\mathbf p}$ the momentum and 
$\hat{n}_{1}$ is the reduced single-particle density function
\begin{eqnarray}
\hat{n}_{1}({\mathbf r},{\mathbf p}|\Gamma) = \sum_{j=1}^{N} 
\delta ({\mathbf r}-{\mathbf r}_{j} ) \, 
\delta ({\mathbf p}-{\mathbf p}_{j} ) \, ,
\label{e9}
\end{eqnarray}
with N being the number of particles, and  $\Gamma$  a point in phase space. 
The corresponding macrovariables are the average of those of Eqs.(\ref{e6}) 
to (\ref{e8}) using the indexed Renyi statistical approach, that is
\begin{eqnarray}
h({\mathbf r},t) = \int d\Gamma \, \hat{h}({\mathbf r}|\Gamma) \: 
{\mathcal D}_{\alpha}(\Gamma,t) \, ,
\label{e10}
\end{eqnarray}	
\begin{eqnarray}
n({\mathbf r},t) = \int d\Gamma \: \hat{n}({\mathbf r}|\Gamma) \: 
{\mathcal D}_{\alpha}(\Gamma,t) \, ,
\label{e11}
\end{eqnarray}	
\begin{eqnarray}
\hat{I}_{n}({\mathbf r},t) = \int d\Gamma \: 
\hat{I}_{n}({\mathbf r}|\Gamma)\: {\mathcal D}_\alpha(\Gamma,t) \, ,
\label{e12}
\end{eqnarray}	

where the integration is over the phase space. Moreover, according to the 
theory \cite{13,21,23,24} it is used the corresponding escort probability 
distribution (cf. Eq.(A11) in Appendix A)
\begin{eqnarray}
{\mathcal D}_\alpha(\Gamma,t) = \varrho_{\alpha}^{\alpha}(\Gamma|t) \Bigg/ 
\int d\Gamma \: \varrho_{\alpha}^{\alpha}(\Gamma|t) \, ,
\label{e13}
\end{eqnarray}	
and it can be noticed that the order of the escort probability is the same 
of the index in Renyi' s distribution \cite{13}, so we keep only one parameter 
in the theory (the so-called infoentropic index $\alpha$), with Renyi's 
heterotypical 
distribution of probability $\varrho _{\alpha}^\alpha(\Gamma|t)$ given 
in Eq.(A13) after using Eq. (A14), both in Appendix A.
 
The equations of evolution for the basic variables associated to the material 
motion (the one of interest for us here) derived in the context of a 
nonlinear kinetic theory \cite{21,42} are
\begin{eqnarray}
\frac{\partial \hfill}{\partial t} n({\mathbf r},t) = 
- \nabla \cdot {\mathbf I}_n({\mathbf r},t) \, ,
\label{e14}
\end{eqnarray}	
\begin{eqnarray}
\frac{\partial \hfill}{\partial t}{\mathbf I}_{n}({\mathbf r},t) = 
- \nabla \cdot I_{n}^{[2]}({\mathbf r},t) + 
{\mathbf J}_{n\alpha}({\mathbf r},t) \, ,
\label{e15}
\end{eqnarray}
with Eq.(\ref{e14}) being the conservation equation for the density, the 
terms with the presence of the divergence operator, $\nabla \, \cdot$, arise 
out of the contribution resulting from	performing the, in this classical case, 
Poisson bracket  with the kinetic energy operator $\hat{H}_{0}$ 
(i.e. $\big\{ \hat{n}, \hat{H}_{0} \big\}$, etc), and 		
\begin{eqnarray}
{\mathbf J}_{n\alpha}({\mathbf r},t) = \lim_{\epsilon \rightarrow +0} 
\int d\Gamma \int_{-\infty}^{t}dt^{'} \, e^{\epsilon(t-t^{'})} \, 
\left\{ \left\{ \hat{{\mathbf I}}_{n}({\mathbf r}|\Gamma) \, , \, 
\hat{H}^{\prime}(\Gamma|(t^{'}-t) \right\} \, \hat{H}^{\prime}(\Gamma) \right\}
\, {\mathcal D}_{\alpha}(\Gamma,t) \, ,
\label{e16}
\end{eqnarray}	
is a scattering integral accounting for the effects of the collisions 
generated by ${\hat H}^{\prime}$, and $I^{[2]}$ is the second-order flux given 
by \cite{21,36,43}
\begin{eqnarray}
I_{n}^{[2]}({\mathbf r},t) = 
\int d^3p \, \frac{[{\mathbf p}{\mathbf p}]}{m^{2}} \, 
\hat{n}_{1}({\mathbf r},{\mathbf p}|\Gamma){\mathcal D}_{\alpha}(\Gamma,t) \, ,
\label{e17}
\end{eqnarray}			
where $[...]$ stands for tensorial product of vectors, rendering a rank-two 
tensor. To solve the system of Eqs. (\ref{e14}) and (\ref{e15}) we need in 
Eq.(\ref{e15}) to express the right-hand side in terms of the basic variables. 
The scattering integral takes in general the form of a relaxation-time 
approach , namely (see Appendix B))
\begin{eqnarray}
{\bf J}_{n\alpha}({\mathbf r},t) = 
- \frac{\hat{{\mathbf I}}_{n}({\mathbf r},t)}{\tau_{I\alpha}} \, ,
\label{e18}
\end{eqnarray}					
where $\tau$ is the momentum relaxation time (see for example Ref.[49]). 
Transforming Fourier in time Eq.(\ref{e15}) we have, after using 
Eq.(\ref{e18}), that
\begin{eqnarray}
\left( 1 + {\textrm i} \, \omega \, \tau_{I\alpha} \right) \, 
{\mathbf I}_{n}({\mathbf r},\omega) = 
- \, \tau_{I\alpha} \nabla \cdot I_{n}^{[2]}({\mathbf r},\omega) \, ,
\label{e19}
\end{eqnarray}
which in the limit of small frequency, meaning 
$\omega \tau_{I\alpha}\ll 1$, reduces to
\begin{eqnarray}
{\mathbf I}_n({\bf r},t) = 
- \, \tau_{I\alpha} \nabla \cdot I_{n}^{[2]}({\mathbf r},t) \, ,
\label{e20}
\end{eqnarray}					
and then, after using Eq.(\ref{e20}) in Eq.(\ref{e15}), we obtain that 	
\begin{eqnarray}
{\frac{\partial \hfill}{\partial t}}n({\mathbf r},t) = 
\tau_{I\alpha} \nabla \cdot \nabla \cdot I_{n}^{[2]}({\mathbf r},t) \, .
\label{e21}
\end{eqnarray}	

In order to close Eq. (\ref{e21}) we need to express the second-order flux 
in terms of the basic variables, {\it n}	and {\bf I}, which after some 
calculus, described in Appendix A, results in that
\begin{eqnarray}
\nabla \cdot \nabla \cdot I_{n}^{[2]}({\mathbf r},t) = 
\xi_{n\alpha} \nabla^{2} n^{\gamma}_{_{\alpha}}({\mathbf r},t) \, ,
\label{e22}
\end{eqnarray}	
with $\xi _{n\alpha}$ given in Eq.(\ref{a23}) and where $\gamma_{\alpha}$ 
is the $\alpha$-dependent fractional power
\begin{eqnarray}
\gamma_{\alpha} = \frac{5-3\alpha}{3-\alpha} \, ,  
\label{e23} 
\end{eqnarray}	
and consequently it follows the so-called "anomalous" diffusion equation
\begin{eqnarray}
\frac{\partial \hfill}{\partial t} n({\mathbf r},t) + 
D_{n\alpha} \nabla^{2} n^{\gamma} _{_{\alpha}}({\mathbf r},t) = 0 \, ,
\label{e24}
\end{eqnarray}	
where $D_{n\alpha} = \xi_{n\alpha} \, \tau_{I\alpha}$ (it can be noticed that 
the dimension of $D$ is cm$^{3 \gamma _{_{\alpha}}-1}$/sec). Moreover, 
the possible values of $\alpha$ having physical meaning 
(i.e. not leading to singularities) belong to the interval [see Appendix A]
\begin{eqnarray}
1 \leq \alpha < \frac{5}{ 3} \, .
\label{e25}
\end{eqnarray}	

At this point we recall, and stress, that the "anomalous" diffusion equation 
is an equation appropriate for approximately describing the  hydrodynamic 
motion, being an artifact (thus the use of the word anomalous between 
quotation marks) of having applied an unconventional statistics in the 
truncated description as given in Eq. (4). As already noticed, in the 
conventional (well established and parameter free) formalism for satisfying 
the principle of sufficiency it must be used a higher-order hydrodynamics.
 
We proceed to analyze this question of "anomalous" results in measurements 
using cyclic voltammetry in fractal-structured electrodes in terms of 
the previous results. Let us, for the sake of completeness, briefly summarize 
the case: The difference of chemical potential between an anode  and a cathode 
with a thin film (nanometric fractal surface) of, say, nickel oxides, produces 
a movement of charges in the electrolyte from the former to the latter. In a 
cyclic voltammetry experiment, these charges circulate as a result of the 
application of a potential $\it{V(t)}$ , with particular characteristics: 
It keeps increasing linearly in time as $\it{V_{0}}+{\it {v}}t$, where 
{\it v} is a scanning velocity, during an interval, say $\Delta$t, and next 
decreases with the same sweep rate, i.e. as 
$V_{0}+{\it v}\bigtriangleup t -{\it v}t$, until recovering the value 
$\it{V_{0}}$. A current ${\it i(t)}$ is produced in the closed circuit, 
which following the potential $\it{V(t)}$ keeps increasing up to a peak 
value ${\it  i_p}$ , and next decreases. This current is the result of the 
movement of the charges that keep arriving to the thin film fractal-like 
cathode. It is found that there follows a power law relation between the 
peak value, ${\it i_p}$, of the current and the rate of change ${\it v}$ of 
the electric field, namely ${\it i_{p}} \sim {\it v^{\epsilon}}$. The value 
of the current to circulate in the cathode, ${\it i(t)}$, is proportional 
to the charge at the interface, that is, the value at it of ${\it n(x,t)}$, 
to be given by the solution of 	Eq.(\ref{e26}) which  in one dimension, 
say, the ${\it x}$-direction normal to the electrode surface, is given by 
\cite{50,51}
\begin{eqnarray}
n(x,t) = b_{\alpha} \, t^{-\mu _{\alpha}} 
\left[ a^{2} + x^{2} \, t^{-2 \mu _{\alpha}} 
\right]^{\frac{1}{\gamma_{_{\alpha}} - 1}} \, ,
\label{e26}
\end{eqnarray}
where ${\it a}$ (which ensures the normalization)and $b_\alpha$ are constants 
of no specific interest in what follows and then we omit to write them down, 
and 
\begin{eqnarray}
\mu _{\alpha} = \left[ \gamma _{\alpha} + 1 \right]^{-1} = 
\frac{1}{4} \left[ \frac{3 - \alpha}{2 - \alpha} \right] \, .
\label{e27}
\end{eqnarray}

Hence the charge at each given point ${\it x}$ at the interface is given by 
Eq.(26), and the current to be produced in  the electrode - motion of this 
charge arriving at it under the action of the field to be applied -  is 
proportional to it, and then we have the law
\begin{eqnarray}
{\textit I}(t) = C_{\alpha} \, t^{-\mu_{\alpha}} \, 
\left[ a^{2} + x^{2} \, t^{-2\mu _{\alpha}} 
\right]_{at \, interface}^{\frac{1}{\gamma_{_{\alpha}} - 1}} \, ,
\label{e28}
\end{eqnarray}		
where $C_{\alpha}$ is a constant.

Taking into account that $\mu _{\alpha}$ is positive (cf. Eqs. (27) and 
(25), and that $0<\gamma _{\alpha}\leq 1$), for not too-short times after 
application of the field we can expect that 
$a^{2}\gg x^{2}t^{-2\mu _{\alpha}}$ and then there follows a power law in 
time for the current, namely
\begin{eqnarray}
{\it i(t)} \approx t^{- \mu _{\alpha}} = 
t^{- \frac{1}{\gamma _{_ {\alpha}} + 1}} = 
t^{- \frac{1}{4} \frac{3 - \alpha}{2 - \alpha}} \, .
\label{e29}
\end{eqnarray}
But taking into account that the applied potential is  
$\it{V}=\it{V_{0}}+{\it {v}}t$ (where  ${\it {v}}$ is the scanning velocity), 
and then  $t=(\it{V}-\it{V_{0}})/{\it {v}}$, Eq. (29) leads to the power law
\begin{eqnarray}
{\textit i}_{p} \approx {\it {v}}^{\epsilon} \, ,
\label{e30}
\end{eqnarray}
where $\epsilon$ is $\mu _{\alpha}$, that is,
\begin{eqnarray}
\epsilon = (\gamma _{\alpha} +1)^{-1} = 
{\frac{1}{4}} \left[ \frac{3-\alpha}{2-\alpha} \right] \, .
\label{e31}
\end{eqnarray}			

In the conditions of the work of Pajkossy and Nyicos \cite{26}, the fractal 
power $\epsilon$ is related to the fractal dimension $d_f$ of the electrode 
rough surface by the relation $\epsilon \simeq (d_f -1)/2$, and then in 
experiments when it holds, from the log-log plot of ${\it i}_{p}$ vs {\it v} 
there follows the value of $\epsilon$ 	and, consequently, the fractal 
dimension can be estimated. Using Eq.(\ref{e31}) we arrive to the relation 
between fractal dimension and the parameter $\alpha$, namely 
$\alpha = (8\epsilon-3)/(4\epsilon -1)$, and then 
\begin{eqnarray}
\alpha = \frac{4d_{f} - 7}{2d_{f} - 3} \, .
\label{e32}
\end{eqnarray}		

We can see that for $d_{f} = 2$ (a perfectly flat surface) we have 
$\alpha = 1$ as it should, while for the other possible extreme limit of 
$d_f = 3$, it follows $\alpha=5/3$. As already noticed, the theory precisely 
restricts the values of $\alpha$ to this interval; outside it the calculations
present singularities. Then as a consequence the exponent $\epsilon$  is 
also limited, and we can only expect for it values in the interval 
$0.5 \leq \epsilon < 1$. Hence, we stress that these are the 
\emph{permited limiting values of the indexes} $\alpha$ \emph{and} $\epsilon$: the 
value which adjusts the experimental data is to be contained between these 
limits, being dependent on the system´s dynamics, geometry and size, 
macroscopic thermodynamic state, and the experimental protocol (in the present 
case depends on the scanning velocity {\it v}, as shall be shown in the next 
section). We can see that the three quantities, $\alpha, \epsilon$ and 
$d_{f}$ are related and give a kind of measure of the microroughness of the 
electrode surface. A perfect one (a totally smooth surface) corresponds to 
$\alpha=1, \epsilon=0.5$ and $d_{f}=2$, and as the surface becomes more and 
more imperfect, these quantities increase, above those values, but with upper 
boundaries, being 1.666 for $\alpha$, 1 for $\epsilon$	and 3 for $d_{f}$. 

On the other hand, atomic force microscopy allows to obtain the fractal 
dimension and a kind of degree of microroughness of the surface. According to 
the method \cite{29}, it is calculated the so-called interface width, 
namely			
\begin{eqnarray}
W(L) = \left[ \frac{1}{L} \sum_{j=1}^{L} 
\left( h_{i} - {\overline h} \right)^{2} \right]^{\frac{1}{2}} \, ,
\label{e33}
\end{eqnarray}	
where L is the number of partitions in which the surface is divided and 
$h_{i} - {\overline h}$ is the deviation, within each partition, of the height 
with respect to the average value. There exists a power law between W and L, 
namely $W \sim  L^{r}$, where r is called the roughness exponent, which is 
related to the fractal dimension as $d_ {f}\simeq 3-r$. Therefore, comparing 
with the theory developed for the case of cyclic voltammetry, we obtain 
relationships relating the different quantities, namely
\begin{eqnarray}
r = 2 \, (1 - {\epsilon} ) \: \: \: \: ; \: \: \: \: 
r = \frac{1}{2} \left[ \frac{5 - 3 \alpha}{2 - \alpha} \right]
\: \: \: \: ; \: \: \: \: 
r = \frac{2 \gamma _{\alpha}}{\gamma_{\alpha} + 1} \, .
\label{e34}
\end{eqnarray}	

Recalling that, according to the theory, $1 \leq \alpha < 5/3$, we have that 
$r$ can only attain the values $0< r \leq 1$, comprising the conditions 
between highest roughness to a perfectly smooth surface, and thus can be 
used as a kind of "order parameter", or, better to say, "degree of 
fractality parameter".
\section{Theory and Experiment} 
We apply the preceding theory to the study of a class of experiments 
\cite{52}. In that work, the fractal dimension of nickel oxide thin films was 
analyzed by means of cyclic voltammetry and compared with the results obtained 
by atomic force microscopy (AFM). 
	  
Figure 1 shows the dependence of the peak current with the sweep-rate 
velocity {\it v}. We can notice that the peak value of the current increases 
continuosly with {\it v}, and also that it can be identified two regions 
where the dependence is very approximately linear (one in the range 
$1\leq v \leq 10$ mV$s^{-1}$ with a more pronounced slope that the other in 
the range $20\leq v$ up to $\sim$ 100 mV$s^{-1}$). Once we do have a log-log 
plot, consequently it can be identified two values of the exponent 
$\epsilon$, in accordance with Eq.(\ref{e30}). Moreover, from equation 
Eq.(\ref{e31}) we obtain that
\begin{eqnarray}
\alpha = \frac{8 \epsilon - 3}{4 \epsilon - 1} \, ,
\label{e35}
\end{eqnarray}
and two values of the infoentropic index ${\alpha}$ can also be determined. 
The calculated values are presented in Table I. The associated statistical 
fractal 
dimension values ($d_{f}$, Eq.(\ref{e32})) are also presented in this table. 
The larger $d_{f}$  values are obtained in the slow-sweep ranges, in comparison 
with those obtained at faster scans. The fact that can be identified 
from this is that the fractal dimension depends on the 
observational scale, which is limited by the experimental protocol, here the 
scanning velocity range: At low values of ${\it v}$, the ions are constrained 
in a spatial region close to the electrode (known as the diffusion layer), 
and are able to sense the fine details of the microroughness at the 
interface. On the other hand, increasing the scan rate also increases the 
length of the diffusion 
layer, and the global morphology of the electrode surface becomes predominant. 
This behavior is particularly expected in the case of sensing by some 
experimental method the surface morphology of the thin film: this surface 
presents grains, and two distinct statistical processes are responsible 
either for the formation of the local surface morphology of individual grains 
or the global thin film surface morphology, formed by the assemblage of all 
grains. The experimental values determined in this work for the infoentropic 
index $\alpha$ are within the expected theoretical range 
($1 \leq \alpha < 5/3$), and the $d_{f}$ values obtained by cyclic 
voltammetry (Table 1) indicate that a stronger asperity can be associated to 
the local grain morphology, in constrast to a smoother thin film global 
morphology.
   
Figure 2 presents a typical AFM micrography. The presence of structured 
grains, with mean size $\approx$180 nm, can be clearly seen. Figure 3 presents 
a log-log plot of the interface width $W(L)$ as a function of the partitions 
L. Two scaling regions (at short and medium L ranges) are evidenced, separated 
by a transition L value of $\approx$140 nm , close to the mean grain size. 
The roughness exponent r values calculated in accordance with Eq.(\ref{e34}), 
and the associated AFM images statistical fractal dimension values are listed 
in Table I. 
The calculated statistical fractal dimensions from both techniques are in 
good agreement. 
Returning to Fig.1, it is worth noticing that the full curve of dependence of 
$i_{p}$ with {\it v}, once we take into account Eqs.(\ref{e30}) and 
(\ref{e31}), can be approximated by the law
\begin{eqnarray}
\alpha(v) \simeq \frac{v + v_{1}}{v + v_{2}} \, ,
\label{e36}
\end{eqnarray}
where $v_{1} = 235\pm 12$ and $v_{2} = 145\pm 8$ mV/s.

It can be noticed that this law in fact reproduces to a good degree of 
approximation (error smaller than 10\%) the two straight lines in the given 
intervals described, and interpreted, above. We can also see that it is 
verified, as it should, a saturating behavior, namely, the infoentropic 
index $\alpha$ tends to the value 1 for large {\it v}, the minimum value it 
can take. Hence, according to Eq.(\ref{e2}) it follow that $d_{f}$ tends 
to 2, what can be interpreted, as already noticed, that as {\it v} increases 
the charges keep "feeling" less and less the microroughness at the interface. 
Hence, for {\it v} very large there follows that $d_{f}\simeq 2$ (an averaged 
smooth surface), while the limit of {\it v} going to zero would evidence 
the more detailed effect of the fractal topography, given in this case the 
values $d_{f} \simeq 2.8$ (then near 3 and then indicating a strong 
roughness) and $\alpha\simeq1.62\pm 0.08$, in accordance with the fact 
that $\alpha$ must remain in the interval as given by Eq.(\ref{e25}), and 
of course $2 \leq d_{f} < 3$.
\section{Concluding Remarks} 
We have considered in detail a quite interesting case consisting of a 
situation present in certain classes of systems where it becomes difficult 
for the researcher to apply the standard formalism in the conventional, well 
established, logically and 
physically sound Boltzmann-Gibbs statistics, because of the presence of 
inhomogeneous fluctuations which are "anomalous" in some sense and eventually, 
as in the case here, involving one type or other of fractality. In that 
cases -- as already noticed -- 
it may be convenient to depart from conventional statistics, introducing 
alternative ones. In  other words, the appropriate use of the Boltzmann-Gibbs 
formalism may in some problems be quite  complicated to apply or, and more 
fundamental, we may be faced 
with an inaccessible characterization constraints (information): 
thus, simplicity or the force 
of circumstances, respectively, would require setting on the use of other 
measures of informational entropies leading to heterotypical probability 
distributions \cite{22} accompanied, as noticed, by the use of their associated 
escort probability \cite{13,23,24,41}, and here we found convenient to use a 
physical statistical mechanics based on Renyi$^{'}$s statistics 
\cite{15,16,23,24}.
   
The matter has been illustrated in this paper by considering its application 
in  a theoretical study of transport in electrodes in microbatteries, 
particularly cyclic voltammetry, which provides information on the 
fractal-like microroughened texture of them, providing an excellent 
illustration for the use and validation of such statistical theory. In 
ideal conditions (quite smooth surface) the peak current in the voltammetry 
experiment is proportional to the square root of the sweep-rate velocity, 
with which the applied field is changing. The standard theory relates this 
result to the diffusion of the charges that are going to reach the thin film 
electrode. But experimental data show that it is verified a power law, with 
a power different than 0.5, and this is referred to as resulting from a 
process of "anomalous" diffusion, as noted in the previous sections. As we 
have noticed, a description in terms of a process of diffusion does not apply, 
once the hydrodynamic motion that proceeds at the interphase 
electrolyte-electrode where is present a morphology with variations (asperity) 
ranging in the nano- to sub-nanometric scale. Hence, a proper description of 
the hydrodynamic motion requires to go well beyond the long-wavelength limit 
of classical (Onsagerian) thermo-hydrodynamics, being required to introduce 
an extended higher-order (sometimes referred-to as mesoscopic) 
thermo-hydrodynamics, together with the introduction of complicated boundary 
conditions.
   
As shown in the Appendix, using Renyi$^{'}$s statistics but ignoring the 
nano- to sub-nanometric-sized roughness of the thin film electrode, it is 
derived an "anomalous" (non-Fickian) diffusion equation, which allows to 
describe the 
experimental results in cyclic voltammetry. The relevant point is 
that the  index $\alpha$, that the Renyi statistics contains, determines the 
power law in the voltammetry experiment which allows to estimate the 
roughness of the electrode surface and, then, its quality and influence on 
the functioning of the device.
	  
In section  III has been presented a theoretical analysis of  results in a 
given experiment  (cyclic voltammetry and atomic force microscopy): in tables 
and figures are shown the results of the study of the index $\alpha$ of Renyi 
statistics, together with other 
power indexes (determined by the former) and the statistical fractal dimension, 
all depending on the experimental  protocol.
	  
Closing this section we notice that the use of an alternative statistics even 
though does not provide a complete physical picture of the problem in hands 
-- once we do have an utter difficulty to handle the appropriate extended 
non-classical 
thermohydrodynamics to be applied -- the sophistication of the formalism allow 
us to obtain a good insight into the physical aspects of the situation. On 
the one hand, as noticed, the use of the escort probability takes care of 
introducing correlations (fluctuations and higher-order variances), and on the 
other hand the heterotypical probability distribution (the one of Renyi here) 
modifies the weight of the Fourier amplitudes $|n(\bf {Q})|$ of the density 
in relation to their classical values in the linear Onsagerian regime 
\cite{47,48}. As previously noticed and 
discussed, classical thermo-hydrodynamics together with Fick's diffusion 
equation is a satisfactory approach while  $|n(\bf {Q})|$ has leading 
contributions for small Q (large wavelengths) and negligible for intermediate 
to large wavenumbers (intermediate to short wavelengths). When not only 
small wavenumbers but ever increasing ones become relevant for the 
description of the motion, i.e. associated to non-negligible amplitudes 
$|n(\bf {Q})|$ contributing in the Fourier analysis of the density, we need to 
introduce a higher-order thermo-hydrodynamics ( with now equations of 
evolution of the Maxwell- Cattano-type, Burnett and super-Burnett-type, and 
so on) \cite{43,44}. In the case we have analyzed, Renyi heterotypical 
distribution, which leads to obtain the "anomalous" diffusion equation of 
Eq.(24), and where $\alpha > 1$, in the restricted hydrodynamic description 
used, decreases what 
would be the values of $|n(Q)|$ in the standard  treatment for short 
wavenumbers, that is, in the domain of classical hydrodynamics, while 
increases those 
amplitudes outside that region, i.e.  in the regime of the higher order 
thermo-hydrodynamics.
\acknowledgments 
We acknowledge finantial support to both Groups from the S\~ao Paulo State 
Research  Foundation (FAPESP). The authors ARV, AG, MUK and RL are National 
Research Council (CNPq) research fellows; TGSC is a FAPESP Pre-Doctoral fellow.
\appendix
\section{''Anomalous'' (non-Fickian) Diffusion}
Let us consider first the conventional case of diffusion, when the principle 
of sufficiency is satisfied, meaning that the several stringent restrictions 
its validity requires are met, namely, local equilibrium, linear Onsager 
relations and symmetry laws, condition of movement with long wavelengths and 
very low frequencies, and weak fluctuations, are verified. A specific 
criterion of validity is given in Ref.[46], where it is considered a 
system composed of two ideal fluids in interaction between them. The 
continuity equation for the flux ${\mathbf I}_n({\mathbf r},t)$ 
[cf. Eq.(19)], after transforming Fourier in time, takes the form:
\begin{eqnarray}
\left( 1 + {\textrm i} \, \omega \, \tau _{n1} \right) 
{\mathbf I}_n({\mathbf r},t) + \tau_{n1} \nabla \cdot 
I_{n}^{(2)}({\mathbf r},t) = 0 \, ,
\label{a1}
\end{eqnarray}
where $\tau _{n1}$ is the momentum relaxation time. But, at very low 
frequencies $\omega\tau _{n1}\ll 1$, and then 
\begin{eqnarray}
{\bf I}_n({\mathbf r},t) \simeq - \tau_{n1} \nabla \cdot 
I^{(2)}({\mathbf r},t) \, ,
\label{a2}
\end{eqnarray}
and a direct calculation tell us that
\begin{eqnarray}
\nabla \cdot I^{[2]}({\mathbf r},t) = \frac{k_ {B} T}{m} 
\bigtriangledown n({\mathbf r},t) \, ,
\label{a3}
\end{eqnarray}
Replacing Eq.(\ref{a3}) in Eq.(\ref{a2}), and the latter in the conservation 
equation for the density, (Eq.(14))  we obtain the usual Fick's diffusion 
equation
\begin{eqnarray}
\frac{\partial \hfill}{\partial t} n({\mathbf r},t) + 
D \nabla^{2} n({\mathbf r},t) = 0 \, ,
\label{a4}
\end{eqnarray}
where $D = {1 \over 3} v_{th}^{2} \, \tau _{n1}$, with 
${1 \over 2} m \, v_{th}^{2} = {3 \over 2} \, k_{B} \, T$; $v_{th}$ is
the thermal velocity and D is the diffusion coefficient, with dimensions 
cm$^{2}$/sec, and $m$ is the mass of each particle.

Let us now go over the unconventional treatment, which is required once one 
is looking forward for an analysis of data on the basis of a description 
in terms of a diffusive movement, when this is not possible, as a consequence 
that diffusion in the microroughened region is governed by not too long 
wavelengths (up to nanometric ones, i.e. $10^{-7}$cm, while the limitation 
of the diffusive region \cite{46} is of the order of $D/ v_{th}$, say, 
typically $10^{-2}$ to $10^{-4}$  cm). A higher-order thermo-hydrodynamics 
\cite{43,44,53} needs be introduced, but if the lower order description 
including only the density and its flux is kept, then sufficiency is not 
satisfied and we need to introduce  auxiliary Statistical Mechanics 
\cite{23,24}. Let us consider the auxiliary statistical operator which for 
this system of identical free particles is in Renyi statistics given by 
\begin{eqnarray}
{\overline\varrho}_{\alpha}(\Gamma|t) = 
\frac{1}{{\overline \eta}_{\alpha}(t)} 
\left[ 1 + (\alpha - 1) \int d^3r \int d^3p \, 
\varphi_{\alpha}({\mathbf r},{\mathbf p};t) \triangle 
\hat{n}_{1}({\mathbf r},{\mathbf p},t|\Gamma) 
\right]^{- \frac{1}{\alpha - 1}} \, ,
\label{a6}
\end{eqnarray}	
where 
\begin{eqnarray}
\triangle \hat{n}_{1}({\mathbf r},{\mathbf p};t) = 
\hat{n}_{1}({\mathbf r},{\mathbf p};|\Gamma) - 
\left\langle \hat{n}_{1}({\mathbf r},{\mathbf p};t) 
\right\rangle_{\alpha} \, ,
\label{a7}
\end{eqnarray}	
with 
\begin{eqnarray}
\left\langle \hat{n}_{1}({\mathbf r},{\mathbf p};t) 
\right\rangle_{\alpha} = \int d\Gamma \, 
\hat{n}_{1}({\mathbf r},{\mathbf p};|\Gamma) \, 
{\mathcal D}_{\alpha}(\Gamma|t) \, ,
\label{a9}
\end{eqnarray}	
and where $ \varphi_{\alpha}({\mathbf r},{\mathbf p},t)$ is the associated 
Lagrange multiplier, $\overline\eta$ ensures its normalization, 
$\hat{n}_{1}$ is given in Eq.(9), and 
\begin{eqnarray}
{\mathcal D}_{\alpha}(\Gamma|t) = 
\left( {\overline\varrho}_{\alpha} \right)^{\alpha} \Bigg/ 
\int d\Gamma \, \left( {\overline\varrho}_{\alpha} 
\right)^{\alpha} \, ,
\label{a11}
\end{eqnarray}	
is the accompanying escort probability (cf. Eq.(1)).

Introducing the modified Lagrange multiplier
\begin{eqnarray}
\tilde{\varphi}_{\alpha}({\mathbf r},{\mathbf p},t) = 
\varphi_{\alpha}({\mathbf r},{\mathbf p},t) \left[ 1 - ( \alpha - 1 )
\int d^{3}r \int d^{3}p \, \, \varphi_{\alpha}({\mathbf r},{\mathbf p},t) 
\left\langle \hat{n}_{1}({\mathbf r},{\mathbf p};t ) 
\right\rangle_{\alpha} \right]^{-1} \, ,
\label{a12}
\end{eqnarray}	
we find that
\begin{eqnarray}
{\overline \varrho}_{\alpha}(\Gamma|t) = 
\frac{1}{{\overline z}(t)} \left[ 
1 + (\alpha - 1) \int d^3 r \int d^3 p \, 
\tilde{\varphi}_{\alpha}({\mathbf r},{\mathbf p},t) \, 
\hat{n}_{1}({\mathbf r},{\mathbf p}|\Gamma) 
\right]^{- \frac{1}{\alpha - 1}} \, ,
\label{a13}
\end{eqnarray}
with ${\overline {z}(t)}$ ensuring the normalization condition. At this point 
we introduce for $\tilde{\varphi}$  the form
\begin{eqnarray}
\tilde{\varphi}_{\alpha}({\mathbf r},{\mathbf p},t) = 
F_{h}({\mathbf r},t) \, \frac{p^{2}}{2 \,m} + F_{n}({\mathbf r},t) +  
{\mathbf F}_{n}({\mathbf r},t) \cdot \frac{{\mathbf p}}{m} \, ,
\label{a14}
\end{eqnarray}	
for, in a such a way, keeping as basic variables the three of Eqs.(10) to 
(12), and thus arriving to the statistical operator of Eq.(13).

Using Eqs. (A.5) to (A.14), after some lengthy but straightforward 
calculations, it follows for the energy density that
\begin{eqnarray}
h({\mathbf r},t) = \int d^3p \, \frac{{\mathbf p}^{2}}{2m} \int d\Gamma \,  
\hat{n}_{1}({\mathbf r},{\mathbf p}|\Gamma) \, 
{\mathcal D}_{\alpha}(\Gamma,t) =
u({\mathbf r},t) + n({\mathbf r},t) \frac{1}{2} \, m \, 
v_{\alpha}^{2}({\mathbf r},t) \, ,
\label{a15}
\end{eqnarray}	
i.e., composed of the energy associated to the drift movement (the last term) 
and the internal energy density
\begin{eqnarray}
u({\mathbf r},t) = \frac{3}{5 - 3 \, \alpha} \, 
\frac{{\mathcal C}_{\alpha}({\mathbf r},t)}{
{\tilde{\beta}}_{\alpha}({\mathbf r},t)} \, 
n^{\gamma _{\alpha}}({\mathbf r},t) \, ,
\label{a16}
\end{eqnarray}	
where
\begin{eqnarray}
{\mathcal C}_{\alpha}({\mathbf r},t) = 
\left\{ \frac{2 \, \pi \, N}{(\alpha - 1)^{3 / 2} \, {\overline z}_{1}(t)} 
\left[ \frac{2 \,m}{\tilde{\beta}_{\alpha}({\mathbf r},t)} 
\right]^{3/2} \, {\mathcal B}(\frac{3}{2},\frac{\alpha}{\alpha - 1} - 
\frac{3}{2}) \right\}^{\frac{2 \, \alpha - 1}{3 - \alpha}} \, .
\label{a17}
\end{eqnarray}	
${\mathcal B}(\nu,x)$ is the Beta function, we have written 
$F_{n \alpha}({\mathbf r},t) = \tilde{\beta}_{\alpha}({\mathbf r},t)$;
${\mathbf F}_{n\alpha}({\mathbf r},t) = m \, 
\tilde{\beta}_{\alpha}({\mathbf r},t) \, {\mathbf v}({\mathbf r},t)$ 
(introducing a "drift velocity" field ${\mathbf v}({\mathbf r},t)$); 
moreover
\begin{eqnarray}
\gamma_{\alpha} = \frac{5 - 3 \, \alpha}{3 \, \alpha - 5} \, ,
\label{a18}
\end{eqnarray}	
and the values of $\alpha$ are restricted to the interval \cite{54}
\begin{eqnarray}
1 \leq \alpha < \frac{5}{3} \, .
\label{a19}
\end{eqnarray}	

Finally, the second order flux is given by
\begin{eqnarray}
I_{n}^{[2]}({\mathbf r},t) &=& \int d^3p \, 
\frac{[ {\mathbf p}{\mathbf p} ]}{m^{2}} \int d\Gamma \, 
\hat{n}_{1}({\mathbf r},{\mathbf p}|\Gamma) \, 
{\mathcal D}_{\alpha}(\Gamma,t) \nonumber \\
&=& n({\mathbf r},t), \left[ {\mathbf v}_{\alpha}({\mathbf r},t) \, 
{\mathbf v}_{\alpha}({\mathbf r},t) \right ] + 
\frac{2}{3 \, m} \, u({\mathbf r},t) \, {\mathbf 1}^{[2]} \, ,
\label{a20}
\end{eqnarray}
where ${\mathbf 1}^{[2]}$ is the unit second- rank tensor, [...] 
stands for the tensorial product of vectors, and it can be noticed that
\begin{eqnarray}
P^{[2]}({\mathbf r},t) = m \, I_{n}^{[2]}({\mathbf r},t) - 
m \, n({\mathbf r},t) \, \left[ {\mathbf v}_{\alpha}({\mathbf r},t) \, 
{\mathbf v}_{\alpha}({\mathbf r},t) \right ] \, ,
\label{a21}
\end{eqnarray}	
is the pressure tensor. Neglecting the terms quadratic in the drift velocity, 
combining the above equations we obtain that
\begin{eqnarray}
I_{n}^{[2]}({\mathbf r},t) = \xi_{n\alpha}({\mathbf r},t) \, 
n^{\gamma_{\alpha}}({\mathbf r},t) \, {\mathbf 1}^{[2]} \, ,
\label{a22}
\end{eqnarray}	
where
\begin{eqnarray}
\xi_{n\alpha}({\mathbf r},t) = \frac{2}{3 \, m} \, 
\frac{5}{5 - 3 \, \alpha} \, \frac{{\mathcal C}_{\alpha}({\mathbf r},t)}{
\tilde{\beta}_{\alpha}({\mathbf r},t)} \, ,
\label{a23}
\end{eqnarray}	
is the quantity present in Eq.(22).
\section{Flux Relaxation}
The scattering integral of Eq.(16) depends, through the statistical operator 
${\cal D}_{\alpha}$, on the non-equilibrium thermodynamic quantities 
$F_{h}$, $F_{n}$, and ${\mathbf F}_{n}$ (related to the Lagrange 
multipliers that the variational method introduces; cf. Appendix A). But  
the latter  are related to the basic variables, $h$, $n$ and 
${\mathbf I}_{n}$, by means of Eqs.(10) to (12), which are considered as being 
equations of state in the associated nonequilibrium thermodynamics 
\cite{21,32,33}. Hence ${\mathbf J}_{n\alpha}({\mathbf r},t)$ is a functional 
of these basic variables. But we do have, on the one hand, that 
$n({\mathbf r}, t) = n_{0} + \Delta n({\mathbf r},t)$ and 
$h({\mathbf r},t) = h_{0} + \Delta h({\mathbf r},t)$, namely their constant 
global values modified by a small variation, that is $\Delta n \ll n_{0}$ 
and, on the other hand, $\mathbf I$ being also small we can write a series 
expansion in it but retaining only the first (linear) contribution
\begin{eqnarray}
{\mathbf J}_{n\alpha}({\mathbf r},t) \simeq \big[ 
\Theta_{\alpha}^{[2]} \big]^{-1} 
\otimes {\mathbf I}_{n}({\mathbf r},t) \, ,
\label{b1}
\end{eqnarray}	
where $\big[ \Theta_{\alpha}^{[2]} \big]^{-1}$ is the second-rank tensor 
(playing the role of the inverse of a tensorial Maxwell-characteristic 
time \cite{32}) of components $\delta J_{i}/\delta I_{j}$, dependent only on 
$n_{0}$ and $h_{0}$. Writing
\begin{eqnarray}
\big[ \Theta_{\alpha}^{[2]} \big]^{-1} = \frac{1}{\tau_{I\alpha}} + 
\big[ \overset{\circ}{\Theta}_{\alpha}^{[2]} \big]^{-1} \, ,
\end{eqnarray}
where
\begin{eqnarray}
\frac{1}{\tau_{I\alpha}} = \frac{1}{3} Tr \left\{ 
\big[ \Theta_{\alpha}^{[2]} \big]^{-1} \right\} \, 
\end{eqnarray}
and the last term on the right of Eq.(B2) being then the traceless part of 
Maxwell characteristic time tensor, neglecting it Eq.(B1) becomes Eq.(18).
\newpage
\newpage
\begin{table}[!hbp]
\vspace{7pt} \centering 
\begin{tabular}{cccc}
{ } & { } & { }Cyclic Voltammetry & \\ \hline
{ } & $\epsilon$ & $\alpha$  & $d_{f}$ \\ \hline    
{ } slow scan rates & { } 0.85 & 1.58  & 2.70 \\ \hline
{ } fast scan rates & { }  0.63 & 1.34 & 2.26   \\ \hline
{ } & { } & { }Atomic Force Microscopy &  \\ \hline
 { } &  & r  & $d_{f}$ \\ \hline  
{ } short L ranges & { }& 0.76 & 2.24  \\ \hline
{ } medium L ranges & { }& 0.27 & 2.73\\ \hline
\end{tabular}
\caption{Exponent $ \epsilon$, Infoentropic index $\alpha$  and fractal 
dimension  $d_{f}$ obtained from cyclic voltammetry  experiments performed 
at slow and fast scan rates. The values of the roughness exponent r and 
fractal dimension  $d_{f}$ obtained from Atomic Force Microscopy images 
at short and medium  L ranges are also presented.}
\label{tab1}
\end{table}
\newpage
\begin{center}
{\bf Figure Captions}
\end{center}
\emph{Figure 1}- Log-log plot of peak current ${\it i}_{p}$ {\it versus} 
sweep rate ${\it v}$.\\
\emph{Figure 2}- AFM micrograph, 1000 nm $\times$ 1000 nm $\times$ 23 nm.\\
\emph{Figure 3} - Log-log plot of interface width $W(L)$ {\it versus} 
partitions L.\\
\newpage
%
%
\begin{figure}[htbp]
\includegraphics[width=14.5cm]{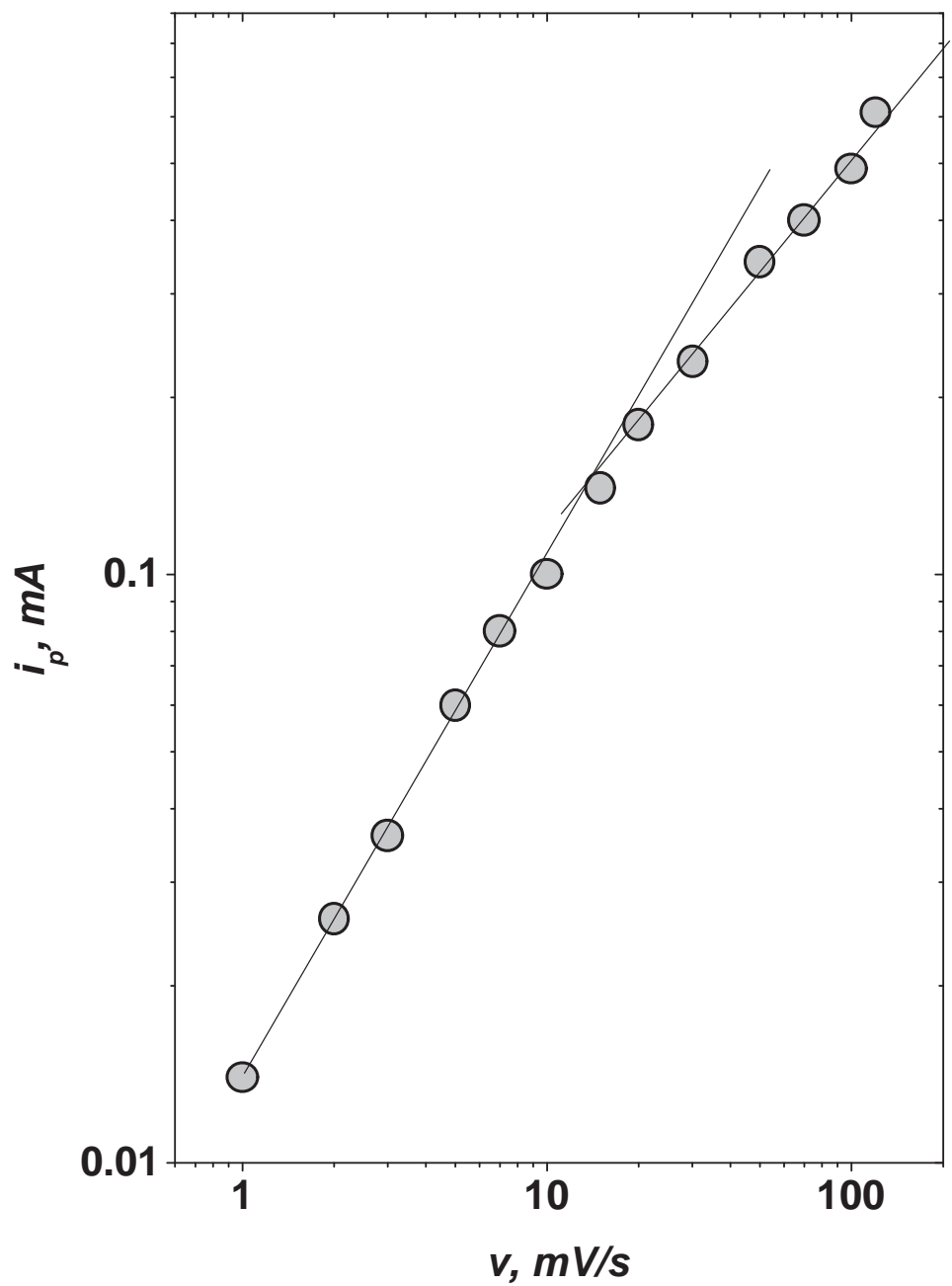}
\caption{Log-log plot of peak current ${\it i}_{p}$ {\it versus} sweep rate 
${\it v}$.}
\end{figure}
\newpage
%
%
\begin{figure}[htbp]
\includegraphics[width=14.5cm]{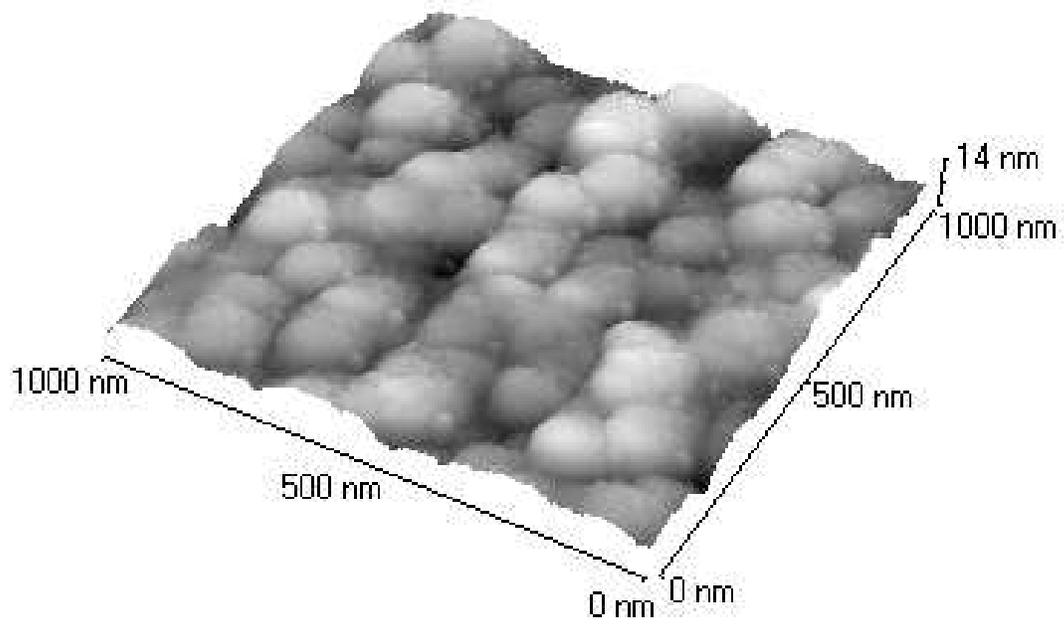}
\caption{AFM micrograph, 1000 nm $\times$ 1000 nm $\times$ 23 nm.}
\end{figure}
\newpage
%
%
\begin{figure}[htbp]
\includegraphics[width=14.5cm]{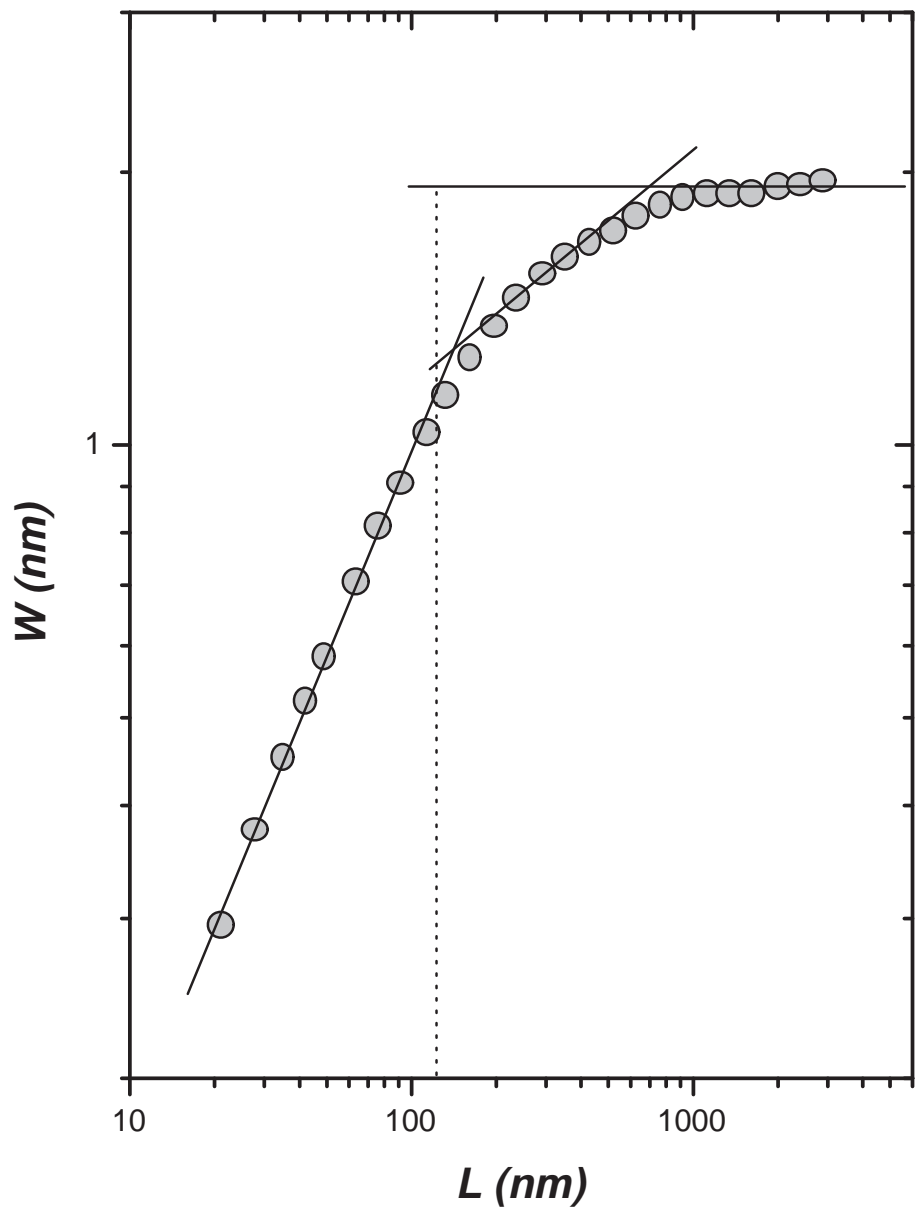}
\caption{Log-log plot of interface width $W(L)$ {\it versus} partitions L.}
\end{figure}
\end{document}